\documentclass[prl,twocolumn,amsmath,amssymb,superscriptaddress]{revtex4-2}
\usepackage{graphicx,amsmath,relsize,epstopdf,color,mathtools,bm,newtxtext,newtxmath,braket,rotating}
\usepackage[hyphenbreaks]{breakurl}
\usepackage[colorlinks=true,linkcolor=blue,citecolor=blue,urlcolor =blue]{hyperref}

\usepackage[table,xcdraw]{xcolor}

\newcommand{\Tr}{\mathop{\mathrm{Tr}} \nolimits}

\begin{document}

\title{Quantumness Beyond Entanglement:  The Case of Symmetric States}

\author{Aaron Z. Goldberg}
\affiliation{National Research Council of Canada, 100 Sussex Drive, Ottawa, Ontario K1A 0R6, Canada}
\affiliation{Department of Physics, University of Toronto, 60 St. George Street, Toronto, Ontario, M5S 1A7, Canada} 

\author{Markus Grassl}
\affiliation{International Centre for Theory of Quantum Technologies, University of Gda\'{n}sk, 80-308 Gda\'{n}sk, Poland}
\affiliation{Max-Planck-Institut f\"{u}r die Physik des Lichts, 91058 Erlangen, Germany}

\author{Gerd Leuchs}
\affiliation{Max-Planck-Institut f\"{u}r die Physik des Lichts, 91058 Erlangen, Germany}
\affiliation{Institute of Optics, Information and Photonics, University of Erlangen-Nuremberg, 91058 Erlangen, Germany}
\affiliation{Institute of Applied Physics, Russian Academy of Sciences, 603950 Nizhny Novgorod, Russia}

\author{Luis~L.~S\'{a}nchez-Soto}
\affiliation{Max-Planck-Institut f\"{u}r die Physik des Lichts, 91058 Erlangen, Germany}
\affiliation{Departamento de \'{O}ptica, Facultad de F\'{\i}sica, Universidad Complutense, 28040 Madrid, Spain}

\begin{abstract}
It is nowadays accepted that truly quantum correlations can exist even in the absence of entanglement. For the case of symmetric states, a physically trivial unitary transformation can alter a quantum state from entangled to separable and vice versa. We propose to certify the presence of quantumness via an average over all physically relevant modal decompositions. We investigate extremal states for  such a measure: SU(2)-coherent states possess the least quantumness whereas the opposite extreme is inhabited by states with maximally spread Majorana constellations.  
\end{abstract}
\maketitle

\emph{Introduction.---}
Entanglement is commonly understood as the inability to describe the state of a compound system in terms of the states of its constituent parts~\cite{Schrodinger:1935ub}. As it stands, this concept may also be applied to different (classical) degrees of freedom of a physical system~\cite{Toppel:2014ug,Aiello:2015tt,Paneru:2020td}. However, the possibility of performing separate measurements on two subsystems, which is a key aspect of entanglement, does not hold for these classical counterparts. Truly quantum entanglement exceeds our understanding of classical correlations~\cite{Laloe:2001wf,Horodecki:2009wo,Brunner:2014vw} and can be ascribed to the intricacies of the measurement process in the quantum domain~\cite{Korolkova:2019vy,Khrennikov:2020wx}. This is the main motivation that fuelled the search for a complete characterization of correlations present in a state~\cite{Ollivier:2001vc,Horodecki:2005vi,Luo:2008vk,Piani:2009ux,Modi:2010tj,Adesso:2010wl,Streltsov:2011tu,Girolami:2014ta,Killoran:2016wz,Adesso:2016uy}.

It is a repeated mantra that entanglement is a fragile, yet crucial resource for performing useful quantum tasks. However, a few \emph{obiter dicta} are in order here. First, the presence of entanglement does not guarantee the quantumness of a state: in polarization optics, the most classical states may be highly entangled~\cite{Goldberg:2020aa,Goldberg:2021tx}. Second, the presence of quantumness does not necessarily require entanglement: there exist separable states that nevertheless exhibit traits without counterpart in the classical world~\cite{Bennett:1999vm}. Third, significantly entangled states need not be fragile: for example, spin-squeezed states are highly entangled, yet particularly robust~\cite{Stockton:2003wf,Leuchs:2005up}.

Photonic systems constitute a particularly versatile platform to implement quantum protocols. But, as optical fields can be decomposed in a variety of fundamental modes (i.e., as it is straightforward to change the partitioning of the Hilbert space into modes), the encoding of quantum information in photons is not unique. Actually, a mode transformation can alter a quantum state from being entangled to being separable and vice versa~\cite{Sperling:2019wg}\footnote{Mode transformations can, in principle, be done with arbitrary quantum states, but they are particularly simple to implement in photonic systems.}.  One might rightly argue that the physics of entanglement in this case should not change just by altering the basis~\cite{Paneru:2020td}, as changing the basis is here akin to producing entanglement by tilting one's head: a waveplate can enact this transformation. In other words,  there is more to quantumness than entanglement: entanglement relies on a preferred decomposition of Hilbert space, whereas quantumness persists in all sensible decompositions. In consequence, a \emph{bona fide} criterion of quantumness should assign no preference to any of these modal decompositions. 

In this paper, we analyze this question for the relevant case of pure two-mode symmetric states, which are permutationally invariant. They can be written as a superposition
\begin{equation}
\ket{\psi} = \sum_{n=0}^{2S} \psi_{n} \,  \ket{n}_a \ket{2S - n}_b \, ,
\label{eq:bipartite pure state}
\end{equation} 
which shows that they contain exactly $2S$ excitations. In this decomposition over modes $a$ and $b$, which may for example correspond to horizontally and vertically polarized states of light, the state is entangled if and only if more than a single coefficient $\psi_{n}$ is nonzero.  The existence of a symmetry simplifies the mathematical description and makes the states experimentally interesting, largely because symmetrically manipulating the system generally requires fewer resources than addressing individual constituents. In particular,  symmetric states are relevant to many experimental situations, such as spin squeezing~\cite{Ma:2011xd}. These states are also numerically tractable, in that the size of their Hilbert spaces grows only linearly with $S$, as opposed to exponentially. For these reasons, there have been numerous attempts to characterize the entanglement properties of symmetric (i.e., bosonlike) states \cite{Chen:2007ut,Devi:2007wy,Ichikawa:2008wi,Toth:2009wu,Kiesel:2010uk,Aulbach:2012wb,Markham:2011wa,Augusiak:2012wy,Wang:2013uw,Daoud:2018vn}. 

A considerable amount of work has since been done using a multipartite description of symmetric states. In this scenario, the Hilbert space of the systems is considered as a tensor product of $2S$ single-qubit Hilbert spaces.  In such a partitioning, changing the modal decomposition as above amounts to a series of local operations and thus does not affect the overall entanglement properties~\cite{Bastin:2009ta,Mathonet:2010tm,Ribeiro:2011ti}.  This has consequently led to  entanglement measures defined from the perspective of a multipartite entanglement~\cite{Aulbach:2010ui,Martin:2010ui,Aulbach:2011ti}. 

Still, it seems natural to address the entanglement properties of states such as \eqref{eq:bipartite pure state}  from a bipartite perspective to properly describe the quantumness found in, e.g., arbitrary spin-$S$ systems. In addition, we should have a mode-independent quantification of the total quantumness present in such a system. In this Letter, we tackle this problem by averaging over all modal decompositions to provide a covariant notion of quantumness. 

Our measure is given by a sum of multipole moments of a state, which allows us to connect quantumness to its geometrical properties. Exploiting the Majorana representation~\cite{Majorana:1932ul,Bengtsson:2006aa}, the problem appears to be closely related to distributing points over the surface of the Bloch (or Poincar{\'e}) sphere. We recall that the question of distributing points uniformly over a sphere has not only inspired mathematical research~\cite{Saff:1997aa,Brauchart:2015aa}, but it has been attracting the attention of physicists working in a variety of fields~\cite{Hannay:1998ab,Makela:2010aa,Lamacraft:2010aa,Bruno:2012aa,Lian:2012aa,Devi:2012aa,Cui:2013aa,Yang:2015aa,Bjork:2015ab,Liu:2016aa,Chryssomalakos:2017aa,Goldberg:2018aa,Chabaud:2020aa}. We find that the most quantum states have these points maximally spread, whereas the most classical states are the SU(2)-coherent states, which are represented by the most concentrated configuration: just a single point. This should prove useful to the many applications in which the quantumness of spin-$S$ states is tied to their advantages in quantum metrology and quantum information protocols.

\emph{SU(2)-covariant measure of bipartite entanglement.---}
To assess the amount of entanglement present in a pure state \eqref{eq:bipartite pure state} we shall use the linear entropy of the reduced density matrices $\varrho_{i}$: $\mathcal{E} (\ket{\psi}) = 1 - \Tr (\varrho_{i}^{2})$ ($i \in \{a, b \}$), 
where $\varrho_{a} = \Tr_{b} (\varrho)$ (analogously for $\varrho_{b}$) and $\varrho = | \psi \rangle \langle \psi |$ is the density matrix of the total system. In terms of the Schmidt coefficients $\psi_{n}$, the linear entropy is given by~\cite{Bengtsson:2006aa} 
\begin{equation}
\label{eq:Schmidtnum}
\mathcal{E} ( \ket{\psi} ) = 1 - \sum_{n=0}^{2S} | \psi_{n} |^{4} \, ,
\end{equation} 
where $\mathcal{E}=0$ implies a separable state and $\mathcal{E}=\frac{2S}{2S+1}$ a fully entangled one, where the former have a single nonzero Schmidt coefficient and the latter, like Bell states, have $2S+1$ Schmidt coefficients with equal magnitude \cite{Tzitrinetal2020}. Since the linear entropy is an entanglement monotone for bipartite pure states, it fully characterizes the entanglement present in this partition. Changing the partition changes the Schmidt coefficients of a state, thereby changing the linear entropy $\mathcal{E}$. 

Transforming the modes is represented by a unitary transformation $ R \in $ SU(2). To make an SU(2)-covariant measure, we average $\mathcal{E}$ over all of the relevant partitions of Hilbert space. Using the normalized Haar measure~\cite{Alfsen:1963wb} $dR$ for SU(2), our averaged entanglement measure reads 
\begin{equation}
\bar{\mathcal{E}} (\ket{\psi}) = \int dR \, \mathcal{E} ( R \ket{\psi}) \, .
\end{equation}
In the language of polarization, this is equivalent to averaging the entanglement found after passing through a random wave plate, thus giving no privilege to a particular basis, such as horizontal and vertical or diagonal and antidiagonal, for analyzing the entanglement. 

The action of $R$ on the coefficients $\psi_{n}$ is not straightforward, so we instead evaluate this quantity resorting to a parametrization of symmetric quantum states that is better suited to describing SU(2) transformations. To this end, we start by expressing the density matrix $\varrho$ as
\begin{equation}
\varrho = \sum_{K=0}^{2S} \sum_{q=-K}^K \varrho_{Kq} \, T_{Kq} \, ,
\end{equation} 
where the irreducible tensors (also called polarization operators) associated with spin-$S$ are given by~\cite{Fano:1959ly,Blum:1981rb}  
\begin{equation}
T_{Kq} = \sqrt{\frac{2 K +1}{2 S +1}} 
 \sum_{m,m^{\prime}=-S}^{S}  C_{Sm, Kq}^{Sm^{\prime}} \,
|  S  m^\prime \rangle \langle S m | \, ,
\end{equation} 
with $C_{S_{1}m_{1}, S_{2} m_{2}}^{Sm}$ denoting the Clebsch-Gordan
coefficients~\cite{Varshalovich:1988ct} that couple a spin $S_{1}$ and a spin $S_{2}$ to a total spin $S$ and vanish unless the usual angular momentum coupling rules are satisfied: $ 0 \leq K \leq 2S$ and $ -K\leq q \leq K$.  These  are $(2S+1)^2$  operators that constitute a basis of the space of linear operators
acting on  the Hilbert space.  The expansion coefficients $\varrho_{Kq} = \Tr ( \varrho \, T_{Kq}^{\dagger})$ are called the state multipoles and contain the complete information about the state sorted in the appropriate way: they are the $K$th-order moments of the generators. Normalization dictates that $\varrho_{00}=1/\sqrt{2S+1}$, and Hermiticity implies $\varrho_{Kq}^{\ast} = (-1)^{q} \varrho_{K -q}$. 

\begin{figure*}
  \centerline{\includegraphics[width=2.1\columnwidth]{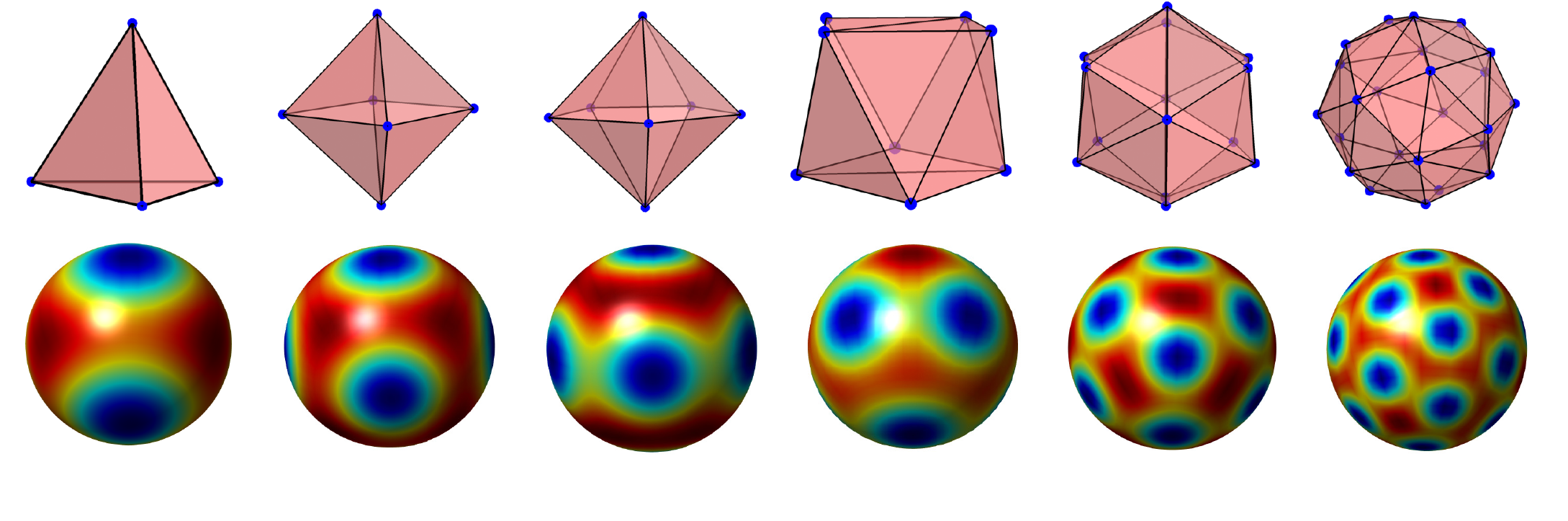}}
  \caption{Density plots of the SU(2) $Q$ functions for
    the most quantum states for the cases $S = 2,  3, 7/2, 4, 6$, and $12$ (from left to right; blue indicates the zero values and red the maximal ones). On top, we sketch  the corresponding Majorana constellation for each state.}
  \label{fig:Qfunc}
\end{figure*}

Due to their very same definition, the multipoles inherit the proper transformation under SU(2); that is, if the state experiences the unitary transformation $\widetilde{\varrho} = R \, \varrho \, R^{\dagger}$, the  multipoles transform as 
  \begin{equation}
 \widetilde{\varrho}_{Kq} =  \sum_{q= -K}^{K}  D_{q^{\prime} q}^{K \ast}(R) \, \varrho_{Kq^{\prime}} \, ,
 \end{equation}
where  $D_{q^{\prime} q}^{K} (R)$ are the Wigner $D$-matrices~\cite{Varshalovich:1988ct}. 

The linear entropy of the transformed state can be computed via the reduced density matrix
 \begin{equation}
 \widetilde{\varrho}_{a} = \sum_{Kq} \sqrt{\frac{2 K +1}{2 S +1}} \;  \widetilde{\varrho}_{Kq}
 \sum_{m=-S}^S C_{Sm,Kq}^{Sm} \, |S m \rangle \langle Sm| \, ,
 \end{equation} 
where only the $q=0$ terms will contribute due to properties of the Clebsch-Gordan coefficients and we abuse the notation $\ket{Sm}$ to denote $\vphantom{a}_b\langle S-m|Sm\rangle=\ket{S+m}_a$. Then, using the orthogonality of the Clebsch-Gordan coefficients, the trace of the square of $\widetilde{\varrho}_a$
 yields
 \begin{equation}
 \mathcal{E} ( R \ket{\psi}) =   1 - \sum_{K,K^\prime} 
 \widetilde{\varrho}_{K0}   \widetilde{\varrho}_{K^\prime 0}^\ast \delta_{K K^\prime} \, .
  \end{equation}
We can then average over the rotations using properties of the $D$-matrices. To this end, we note that
 \begin{align}
  \int dR \, \widetilde{\varrho}_{K0} \widetilde{\varrho}_{K^\prime 0}^\ast & =  
 \sum_{q, q^\prime}  \varrho_{Kq} \varrho_{K^\prime q^\prime} ^\ast
 \int dR \, D_{q0}^{K\ast} \, D_{q^\prime 0}^{K^\prime} \nonumber \\ 
 & = \frac{\delta_{KK^\prime}}{2K+1}\sum_{q=-K}^{K} |\varrho_{Kq} |^2.
 \end{align}
 The  averaged entanglement thus becomes 
 \begin{equation}
 \bar{\mathcal{E}} (\ket{\psi}) = 1 -\sum_{K=0}^{2S}\frac{1}{2K+1}\sum_{q=-K}^K \lvert \varrho_{Kq} \rvert^2 \, ,
 \label{eq:averaged entanglement}
 \end{equation}
providing a simple metric for analyzing the quantumness. As $\bar{\mathcal{E}} (\ket{\psi})$ involves all the multipoles, it provides a complete characterization of the state. For the case of pure states we are dealing with,  we expand in the angular-momentum basis as $\ket{\psi} = \sum_{m} \psi_{m} \ket{S,m}$, so \eqref{eq:averaged entanglement} takes the  form
\begin{equation}
 \bar{\mathcal{E}} (\ket{\psi} )= 1 - \frac{1}{2S+1} \sum_{K=0}^{2S} \sum_{q=-K}^K \left | \sum_{m,m^{\prime}=-S}^{S} C_{Sm,Kq}^{Sm^{\prime}} \psi_{m^{\prime}} \psi_{m}^{\ast} \right |^2 \, ,
 \label{eq:averaged entanglemen pure}
 \end{equation}
 which is the quantumness measure we advocate.

\emph{Extremal states.---}
The averaged linear entropy \eqref{eq:averaged entanglemen pure} can be regarded as a nonlinear functional of the density matrix. The higher the value of $\bar{\mathcal{E}}$, the greater the value of the average entanglement. Some pure states give the maximal value of $\mathcal{E}$ for a given partition but no pure state achieves $\mathcal{E}=\frac{2S}{2S+1}$ for all partitions. Maximally mixed states, in contrast, give the maximum value of $\bar{\mathcal{E}}$, but linear entropy is only an entanglement measure for pure states. For this reason, we will restrict our investigation to pure states, ¡¡¡

We first try to ascertain states that minimize $\bar{\mathcal{E}}$. In Ref.~\cite{Bjork:2015ux}, it was claimed that the cumulative multipolar distribution $\mathcal{A}_M \equiv \sum_{K=0}^{M}\sum_{q=-K}^K \lvert \varrho_{Kq} \rvert ^2$ is maximal for SU(2)-coherent states for all $M\leq S$. These states are defined as~\cite{Perelomov:1986ly}  $ \ket{\theta, \phi} = (1 +  |\alpha|^{2})^{-S} \exp(\alpha {S}_{+} ) \ket{S, -S}$, where ${S}_{\pm} = {S}_{x} \pm i {S}_{y}$ are the ladder operators for SU(2) and the complex number $\alpha$ corresponds to the stereographic projection of the point $(\theta, \phi)$ on the sphere; viz, $\alpha = \tan (\theta/2) e^{-i \phi}$.  The monotonicity of the coefficients $1/(K+1)$ immediately implies that the SU(2)-coherent states minimize $\bar{\mathcal{E}}$, with a value
\begin{equation}
1- \bar{\mathcal{E}}_{\mathrm{coh}} = \frac{1}{4S+1}
\sum_{m=0}^{2S} \binom{2S}{m}^2 \binom{4S}{2m}^{-1} =
\frac{\sqrt{\pi} \Gamma(2S+1)}{2 \Gamma(2S+3/2)} \, .
\end{equation}
This accords with many other quantumness indicators agreeing that SU(2)-coherent states, which correspond to a single point on the surface of the sphere, are the least quantum~\cite{Goldberg:2020aa}. 

Next, we concentrate on maximizing $\bar{\mathcal{E}}$. If we write the set of unknown normalized state amplitudes in Eq.~\eqref{eq:averaged entanglemen pure}  as $\psi_{m} = a_{m}+i b_{m}$ ($a_{m}, b_{m} \in \mathbb{R}$), finding the maxima corresponds to a (quartic) polynomial program~\cite{Laserre:2015uw} that can be solved by standard methods. We provide a complete list of the numerical solutions found for different values of $S$ up to 15 in Ref.~\cite{Markus}.

Although the coefficients $\psi_{m}$ completely characterize $\ket{\psi}$, they do not provide a lucid picture of the state. To this end, we will use the concept of Majorana representation~\cite{Majorana:1932ul,Bengtsson:2006aa}, which maps every $(2S+1)$-dimensional pure state $\ket{\psi}$ into the polynomial
\begin{equation}
  \label{eq:MajPol}
  \psi ( \theta, \phi) = \langle \theta, \phi | \psi \rangle \propto  \sum_{m=-S}^{S} \sqrt{\frac{(2S)!}{(S-m)! (S+m)!}}
  \psi_{m} \, \alpha^{S+m} \, .
\end{equation}
Up to a global unphysical factor, $\ket{\psi}$ is determined by the set $\{ \alpha_{i} \}$ of the $2S$ complex zeros of $\psi (\theta, \phi)$, suitably completed by points at infinity if the degree of $\psi (\theta, \phi) $ is less than $2S$.  A nice geometrical representation of $\ket{\psi}$ by $2S$ points on the unit sphere (often called the constellation) is obtained by an inverse stereographic map of $\{ \alpha_{i} \}$. Two states with the  same constellation are the same, up to a global phase. For example, the SU(2)-coherent states have all $2S$ of the ``stars'' in their constellation co-located at angular coordinates $(\theta,\phi)$. Several decades after its conception, this stellar representation has recently attracted a great deal of attention in several fields~\cite{Hannay:1998ab,Makela:2010aa,Lamacraft:2010aa,Bruno:2012aa,Lian:2012aa,Devi:2012aa,Cui:2013aa,Yang:2015aa,Bjork:2015ab,Liu:2016aa,Chryssomalakos:2017aa,Goldberg:2018aa,Chabaud:2020aa}.

Intimately related to the Majorana polynomial $\psi (\theta, \phi)$ is the SU(2) $Q$-function, defined as $Q(\theta, \phi) = \lvert \phi (\theta, \phi) \rvert^{2}$. Obviously, the stars $\{ \alpha_{i} \}$ are also the zeros of $Q (\theta, \phi)$, so the $Q$-function is an attractive way of depicting the state to help appreciate the symmetries of $|\psi \rangle$. It is not surprising that it has gained popularity in modern quantum information~\cite{Goldberg:2020aa}.

\begin{table}[t]
  \caption{Symmetries of the constellations associated to the maximal states for the values of $S$.}
  \medskip
  \label{table1}
  \begin{ruledtabular}
    \begin{tabular}{lllc}
      $S$ & Group &  Order & Constellation  \\
      \hline
 1 & $C_{2}$ & 2  &  --- \\
 $\tfrac{3}{2}$ & $S_{3}$ & 6  & triangle\\
 2 & $S_{4}$ & 24  & Platonic \\
 $\tfrac{5}{2}$ & $D_{12}$ & 12 & triangle + poles     \\
 3 & $C_{2} \times S_{4}$ & 48  & Platonic \\
 $\tfrac{7}{2}$ & $D_{20}$ & 20  & pentagon + poles \\
 4 & $D_{16}$ & 16  & twisted cube  \\
 5 & $D_{16}$ & 16  & twisted cube + poles\\
 6 & $C_{2} \times A_{5}$ & 120  & Platonic  \\
 7 & $D_{24}$ & 24  & twisted hexagon + poles  \\
 8 & $A_{4}$ & 12  & --- \\
 12 & $S_{4}$ & 24  & --- 
    \end{tabular}
  \end{ruledtabular}
\end{table}

The $Q$-functions and the corresponding Majorana constellations for a few examples of extremal states are shown in Fig.~\ref{fig:Qfunc}, with many more given in Ref. \cite{Markus}. The resulting constellations have the points symmetrically placed on the unit sphere, which agrees with other previous notions of quantumness, such as states of maximal Wehrl-Lieb entropy~\cite{Baecklund:2014ng}. 

In particular dimensions, the constellations show a remarkable additional degree of symmetry, some of which are summarized in Table~\ref{table1}. Most of the constellations corresponding to Platonic solids appear in the table and they correspond to states with maximal average entanglement but, surprisingly, states whose constellations correspond to a twisted cube have higher average entanglement than those corresponding to a cube. These states have amazing features: they are maximally unpolarized~\cite{Bjork:2015ux}, they have the highest sensitivity to small misalignment of Cartesian reference frames~\cite{Kolenderski:2008mo}, and they  are optimal to estimate rotations about any axis~\cite{Bouchard:2017aa}. 

Other criteria of quantumness have been considered in this context. Among them, the Kings~\cite{Bjork:2015ux} and the Queens~\cite{Giraud:2010db} of Quantumness seem to be closely related to our approach. For some dimensions, the optimal states turn to be the same, but for others, they are different~\cite{Goldberg:2020aa}, highlighting the rich physics underlying symmetric states and sphere point picking.

\begin{figure}[b]
  \centerline{\includegraphics[width=0.95\columnwidth]{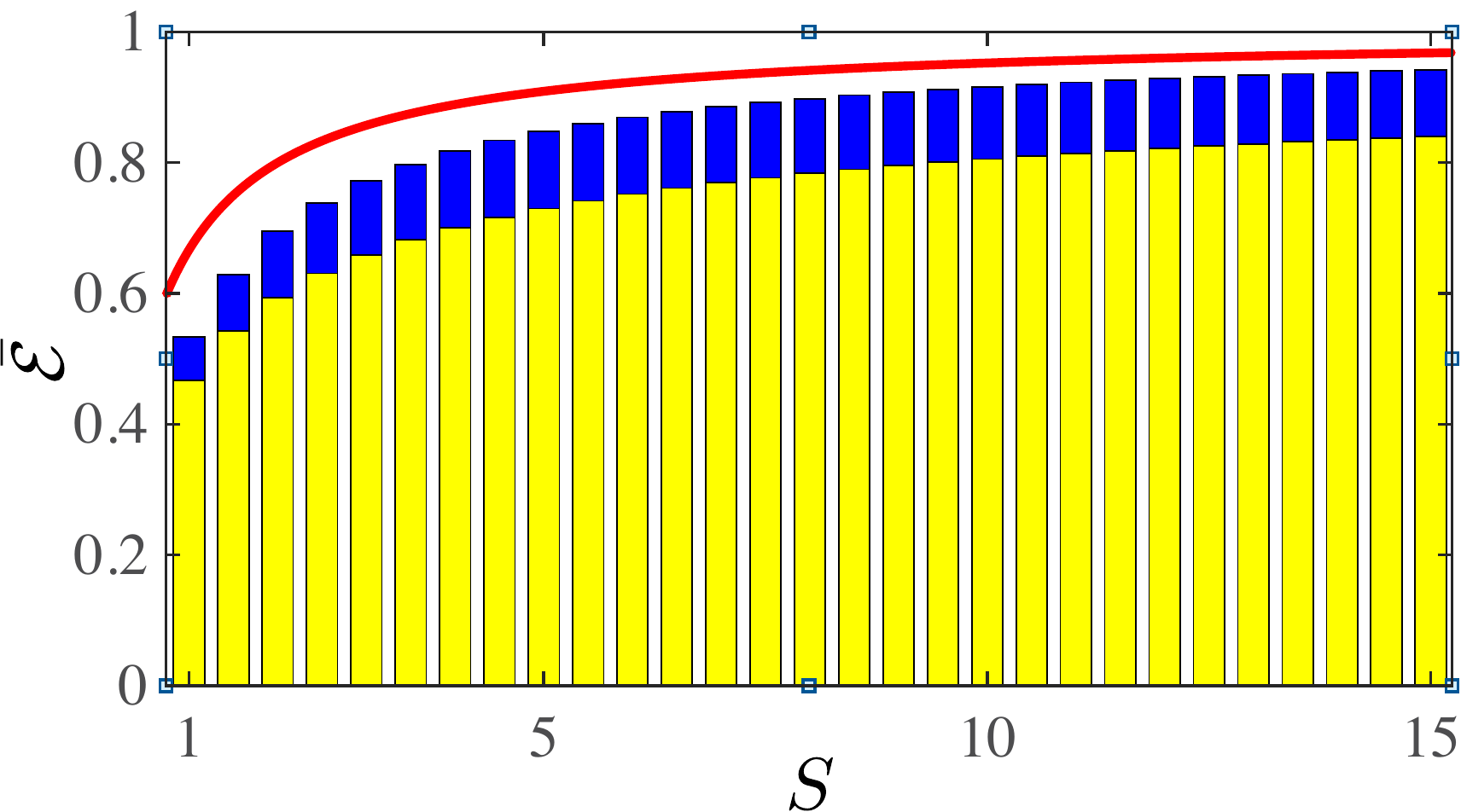}}
  \caption{Average entanglement $\bar{\mathcal{E}}$ for the states of maximal  (blue bars) and of minimal average entanglement (yellow bars) as a function of $S$. The continuous red line represents the upper limit $\mathcal{E} = \frac{2S}{2S+1}$ attainable in \eqref{eq:Schmidtnum}.}
  \label{fig:Emax}
\end{figure}

In Fig.~\ref{fig:Emax} we plot the value of the averaged entropy $\bar{\mathcal{E}}$ for the maximal states found numerically as a function of the dimension $S$. For comparison, we have also included the corresponding values for the minimal states, which correspond to coherent states. As we can appreciate,  $\bar{\mathcal{E}}$ approaches the limit value of unity as $S$ grows. One can easily guess that $\bar{\mathcal{E}} \sim 1 - 1/(2S)$, which shows that the higher the value of $S$, the more quantum the extremal state is.

\emph{Concluding remarks.---}  
We have comprehensively examined the notion of average entanglement for symmetric states, which is the physically relevant quantity for these states. We have proven that SU(2) coherent states are minimal. Their opposite counterparts, maximizing the average entanglement, have interesting properties.  Apart from their indisputable geometrical beauty, there surely is plenty of room for the application of these states.

\emph{Acknowledgments.---} We are indebted to G. Bj\"{o}rk,  P. de la Hoz, J. L. Romero, and K.  \.{Z}yczkowski for discussions.   We acknowledge financial support from the European Union Horizon 2020 (Grants ApresSF and Stormytune), the Ministry of Education and Science of the Russian Federation (Mega-Grant No.14.W03.31.0032), and the Spanish MINECO (Grant PGC2018-099183-B-I00). AZG acknowledges funding from an NSERC Discovery Award Fund, an NSERC Alexander Graham Bell Scholarship, the Walter C. Sumner Foundation, the Lachlan Gilchrist Fellowship Fund, a Michael Smith Foreign Study Supplement, and Mitacs Globalink. The `International Centre for Theory of Quantum Technologies' project (contract No. MAB/2018/5) is carried out within the International Research Agendas Programme of the Foundation for Polish Science co-financed by the European Union from the funds of the Smart Growth Operational Programme, axis IV: Increasing the research potential (Measure 4.3).

%

\end{document}